\documentstyle[epsfig]{mn}
\begin{document}
\def\ltsima{$\; \buildrel < \over \sim \;$}
\def\simlt{\lower.5ex\hbox{\ltsima}}
\def\gtsima{$\; \buildrel > \over \sim \;$}
\def\simgt{\lower.5ex\hbox{\gtsima}}
\def\approxgt{\mathrel{\hbox{\rlap{\lower.55ex \hbox {$\sim$}}
        \kern-.3em \raise.4ex \hbox{$>$}}}}
\def\approxlt{\mathrel{\hbox{\rlap{\lower.55ex \hbox {$\sim$}}
        \kern-.3em \raise.4ex \hbox{$<$}}}}

\title{INTEGRAL discovery of unusually long broad-band X-ray activity from the Supergiant Fast X-ray Transient  IGR~J18483$-$0311}

\author[Sguera et al.]
{V. Sguera$^{1}$, L. Sidoli$^{2}$, A.J. Bird$^{3}$, A. Bazzano$^{4}$\\
$^1$ INAF, Istituto di Astrofisica Spaziale e Fisica Cosmica, Via Gobetti 101, I-40129 Bologna, Italy \\
$^2$ INAF, Istituto di Astrofisica Spaziale e Fisica Cosmica, Via E. Bassini  15, I-20133 Milano, Italy \\
$^3$ School of Physics and Astronomy, Faculty of Physical Sciences and Engineering, University Southampton, SO17 1BJ, UK \\
$^4$ INAF, Istituto di Astrofisica  e Planetologia Spaziali, Via Fosso del Cavaliere 100, I-00133 Rome, Italy\\
}

\date{Accepted 2015 February 16.  In original form 2014 December 16}

\maketitle
{}
\title[A long outburst from the SFXT IGRJ18483-0311 observed by INTEGRAL]{}

\begin{abstract}
We report on a broad-band X-ray study  (0.5--250 keV) of the Supergiant Fast X-ray Transient  IGR J18483$-$0311 
using archival INTEGRAL data    and a new targeted  {\it XMM-Newton} observation.  
Our INTEGRAL investigation discovered for the first time an unusually long X-ray activity (3--60 keV) which continuously lasted for  
at least $\sim$ 11 days,  i.e.  a  significant fraction ($\sim$ 60\%) of the entire orbital period,  and spanned  orbital phases corresponding to both periastron and   apastron passages.  This  prolongated X-ray activity is at odds with the much shorter durations  marking outbursts  
from classical SFXTs especially above 20 keV, as such it represents a departure from their nominal behavior  and it adds a further extreme  characteristic to the already extreme SFXT  IGR~J18483$-$0311.  Our  IBIS/ISGRI high energy investigation (100--250 keV)  of archival outbursts  activity from  the source showed that the recently reported  hint of a possible hard X-ray tail is not real and  it is likely due  to noisy background. The new {\it XMM-Newton} targeted observation did not  detect any sign of  strong X-ray outburst  activity from the source despite being performed close to its periastron passage, on the contrary IGR J18483$-$0311 was caught during the common intermediate X-ray state with  a low luminosity value of $\sim$ 3$\times$10$^{33}$ erg   s$^{-1}$ (0.5--10 keV).  We discuss all the reported results in the framework of both spherically symmetric clumpy wind scenario and   quasi-spherical settling accretion model.
\end{abstract}

\begin{keywords}
X-rays: binaries -- X-rays: individual IGR J18483$-$0311
\end{keywords}

\vspace{1.0cm}

\maketitle

 \section{INTRODUCTION}
The IBIS/ISGRI  instrument onboard the  {\it INTEGRAL} observatory, launched in October 2002,  
has inaugurated  a new era in the study of  High Mass X-ray Binaries (HMXBs) by discovering a peculiar sub-class during  systematic scans of the Galactic plane: the  Supergiant  Fast X-ray Transients (SFXTs).  SFXTs usually host a  neutron star compact object orbiting around a 
massive and hot OB blue supergiant star  as  companion donor (Negueruela et al. 2006),   
in the  X-ray band they display a peculiar fast X-ray transient behavior lasting  
typically from a few hours to no longer than a  few days (Sguera et al. 2005, 2006). Notably the typical  dynamic ranges, from X-ray outbursts (L$_x$  $\sim$ 10$^{36}$ $-$ 10$^{37}$ erg s$^{-1}$) to the lowest level of X-ray emission,  are of the order of  10$^3$--10$^5$:  SFXTs are  among  the most extreme and exotic  X-ray transients in our Galaxy.  This exceptional fast X-ray transient   behaviour   is at odds  with the  bright and persistent 
X-ray emission (L$_x$ $\sim$10$^{36}$erg s$^{-1}$ ) characterizing  their  historical parent population of wind-fed  
Supergiant High Mass X-ray Binaries (SGXBs). 

Despite the elusive nature which makes their discovery challenging,  during the last  nine years  $\sim$ 10 firm SFXTs have been 
reported in the literature (see list in Paizis \& Sidoli 2014) plus a similar number of candidates (e.g. Sguera et al. 2013). 
SFXTs could represent a significant portion of the entire population of  HMXBs in our Galaxy, in particular they could constitute a large fraction of HMXBs with supergiant companions (Ducci et al. 2014). 


The physical mechanism driving the peculiar X-ray behaviour of SFXTs is far from solved, although several  different models have been proposed in the last decade.
All invoke  a structured and inhomogeneous  supergiant stellar wind (the so-called clumpy wind model) and could be broadly divided into two different groups: i)  models involving a neutron  star  compact object with standard magnetic field values (e.g. 10$^{11-12}$ Gauss)  acting as a probe of the clumped supergiant wind  properties;  the fast X-ray outbursts are assumed to be due to accretion  of single and very massive clumps embedded in a much less dense background 
wind (Walter \& Zurita 2007, Sidoli et al. 2007, Negueruela et al. 2008, Ducci et al. 2009),  ii) models involving a slowly rotating 
(e.g. $\ge$ 1,000 s) and highly magnetized neutron star (e.g. $\sim$ 10$^{14-15}$  Gauss)  where  the accretion of single massive clumps 
is prevented  or allowed by the switching of a magnetic or centrifugal barrier, the conditions for their onset can be modified  by even  modest variations of the clumpy wind  parameters (e.g. density and speed)  as expected for example  close to the periastron passage  (Bozzo et al. 2008, Grebenev \& Sunyaev 2007). All such models rapidly became the preferred mechanisms to explain the overall behavior of SFXTs since the early days of their discovery. However in  recent years new observational findings clearly showed that these models  have several shortcomings (e.g. Oskinova et al. 2012), and can hardly explain  the different X-ray  behavior between SFXTs and their historical parent population of persistent SGXBs and are not able to justify the whole observational phenomenology  of  SFXTs. In this respect,  an interesting step forward has been very recently represented by the novel theoretical development proposed  by  Shakura et al. (2012, 2013):  the model of quasi-spherical settling accretion. Although it  was initially  proposed  to explain the off states observed in slowly rotating  X-ray pulsars hosted in HMXBs,  it has since been applied to the case of SFXTs (Drave et al. 2013, 2014,   Paizis \& Sidoli 2014, Shakura et al. 2014).  Such a model, which has also the benefit of elegantly explaining the different X-ray behavior  between the persistent SGXBs and the SFXTs,  is not  mutually exclusive from the clumpy wind scenario but  it does not have its main shortcomings and limitations. In this model  the  accretion onto the neutron star compact object is mediated through a quasi-static shell of plasma above the magnetosphere, the actual mass accretion rate depends on the ability of the plasma shell to enter the neutron star magnetosphere which in turn depends on the cooling mechanism. The fast flaring events observed from SFXTs  are likely generated by a transition from a radiatively inefficient  cooling regime (which allows only very low accretion rate onto the neutron star)   to a much more efficient Compton cooling regime (which allows for a brief period a much higher accretion rate). This transition occurs when the X-ray luminosity decreases/increases below/above the critical value of $\sim$  3$\times$10$^{35}$ erg s$^{-1}$ (Shakura et al. 2012, 2013).

IGR~J18483$-$0311 is among the most studied  SFXTs. It  was discovered in outburst with INTEGRAL  
in 2003 (Chernyakova et al. 2003) during observations of the Galactic center region. Since then several other  X-ray outbursts were observed by INTEGRAL showing hard X-ray fluxes up to  $\sim$ 10$^{-9}$erg cm$^{-2}$   s$^{-1}$
and typical durations from $\sim$ 0.2  to 3 days (Sguera et al. 2007, Ducci et al. 2013). The optical counterpart  has been identified with a massive  B0.5-B1 supergiant star located at a distance  in the range 2.8-4 kpc (Rahoui \& Chaty 2008, Torrejon et al. 2010). The X-ray emission shows two periodicities: the  longer one (at $\sim$ 18.5 days)  is interpreted as  due to the orbital period (Levine \& Corbet 2006, Sguera et al. 2007) while the  shorter one (at $\sim$ 21 seconds)   was discovered
with the soft X-ray monitor JEM-X onboard INTEGRAL (Sguera et al. 2007) and it is interpreted as due to the spin of the neutron 
star compact object.  Giunta et al. (2009) confirmed the  detection of X-ray  pulsations through {\it XMM-Newton} data\footnote{This spin period has been recently called into question by Ducci et al. (2013)}. The lowest X-ray luminosity states  have been measured  at  $\sim$ 1.3$\times$10$^{33}$erg s$^{-1}$ in the soft X-rays (Sguera et al. 2010, Giunta et al. 2009) and  at $\sim$ 1.3$\times$10$^{34}$erg s$^{-1}$ in the 
hard X-rays (Sguera et al. 2010), by assuming a distance of 3 kpc. Monitoring of the soft X-ray emission over an entire orbital period has been performed in 2009 with Swift/XRT (Romano et al. 2010).  Sguera et al. (2010) reported on a possible cyclotron feature  in the {\it XMM-Newton}
spectrum which would imply a neutron star magnetic field of the order of  3$\times$10$^{11}$ Gauss. From a broad band X-ray spectral study of  IGR~J18483$-$0311 in outburst, Ducci et al. (2013) reported the hint of  a  possible hard X-ray tail at energies above $\sim$ 80 keV and extending up to $\sim$ 250 keV. \\
Here we report new results on the SFXT IGR~J18483$-$0311  both below and above 10 keV as obtained  from  a new  targeted {\it XMM-Newton}
observation and archival INTEGRAL  data, respectively.

\begin{figure*}
\begin{center}
\includegraphics[width=.82\textwidth]{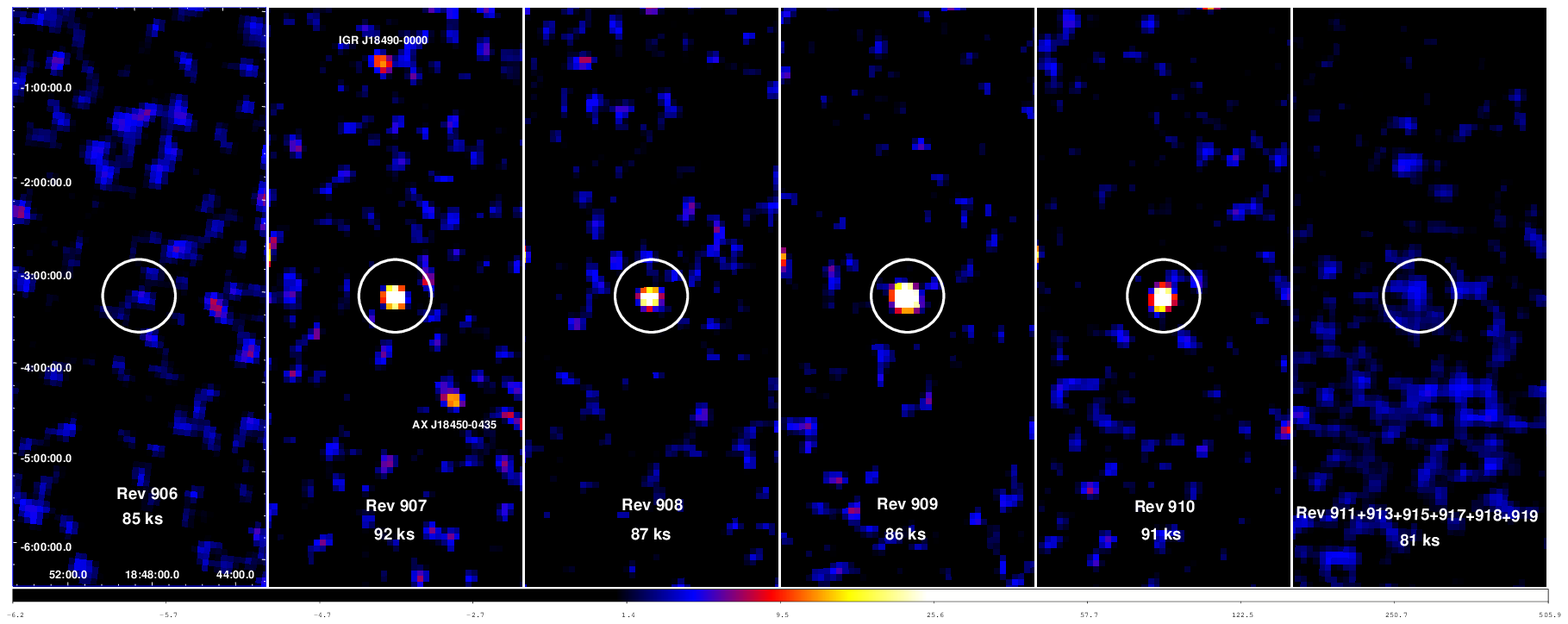}
\includegraphics[width=.82\textwidth]{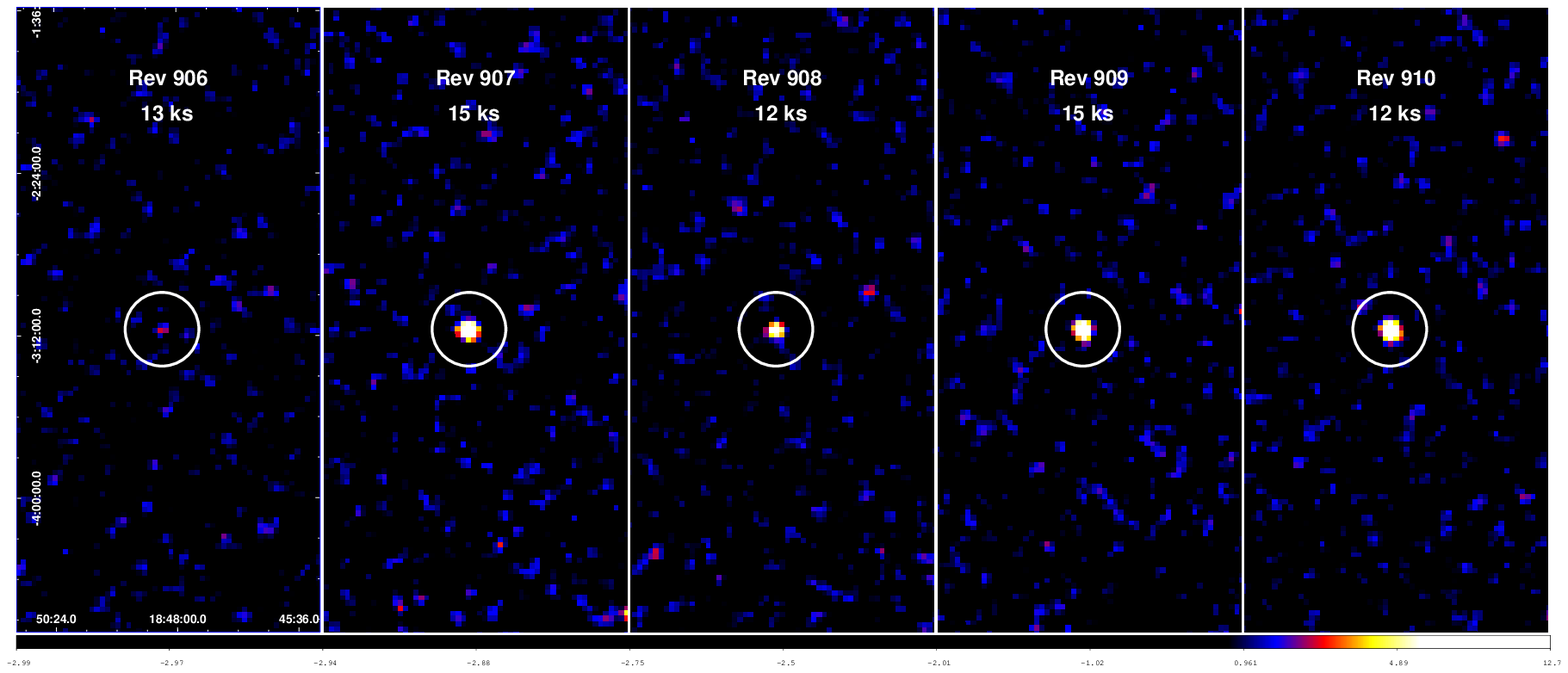}
\caption{IBIS/ISGRI revolution image sequence (18--60 keV, top) and INTEGRAL/JEM-X revolution image sequence (3--10 keV, bottom) 
of the outburst X-ray activity detected from IGR~J18483$-$0311 (encircled).} 
\end{center}
\end{figure*} 

\begin{table*}
\begin{center}
\caption {Characteristics of the X-ray outburst from IGR~J18483$-$0311 analyzed in this paper. Average fluxes and significance values  are provided 
in the energy band 18--60 keV for IBIS/ISGRI and 3--10 keV for JEM--X.} 
\label{tab:main_outbursts} 
\begin{tabular}{cccccccc}
\hline
\hline   
orbital           &   IBIS exp     & IBIS sig                                & IBIS flux             &   JEM-X exp       & JEM-X sig               & JEM-X flux     \\
revolution    &     (ks)           &                                            &  (mCrab)             &  (ks)                     &                                      & (mCrab)     \\
  \hline    
906              &      85              &                                     &   $<$ 0.9$\dagger$                 &    13                   &     2.5$\sigma$      &  2.3               \\         
907                &      92              &     15.6$\sigma$            &  7.3                           &    15                   &     11$\sigma$          &    10.3             \\      
908             &      87              &      13.0$\sigma$             &     6.8                            &    12                   &     8$\sigma$                  &  9.2               \\   
909               &      86              &     24.5$\sigma$             & 11.3                           &    15                   &     12$\sigma$              &    11.2             \\   
910            &      91             &       22.0$\sigma$               &    10.6                            &    12                   &     11$\sigma$               &   11.3              \\   
911-919       &      81              &                                    &  $<$ 1$\dagger$         &    0                     &                                       &                             \\   
\hline
$\dagger$ = 2$\sigma$ upper limit \\ 
\end{tabular}
\end{center}
\end{table*}

\section{DATA ANALYSIS}
\subsection{INTEGRAL}  

We have used data collected with the gamma-ray imager   IBIS/ISGRI (Ubertini et al. 2003, Lebrun et al. 2003) onboard the INTEGRAL satellite (Winkler et al. 2003)
from the end of 2003 February to the end of 2010 December. In particular, the  IBIS/ISGRI  data set consists of   4735 pointings or Science Windows 
(ScWs, typical duration 2000 seconds) where IGR~J18483$-$0311 was within 12$^\circ$    from the centre of the IBIS/ISGRI  fiel d of view (FoV) with an exposure  greater than at least 500 seconds.   We applied a  12$^\circ$ limit  because the off-axis response of IBIS/ISGRI is not well modelled at large off-axis angles and in combination with the telescope dithering (or the movement of the source within the FoV) it may introduce  systematic errors in the measurement of the source fluxes.   IBIS/ISGRI images for each ScW were generated in the energy band 18--60 keV and  count rates at the position of the source were then extracted from all  individual images to produce the 18--60 keV long-term light  curve on the ScW time scale. 

The X-ray monitor JEM--X  (Lund et al. 2003)  on board the {\it INTEGRAL} has a $\sim$ 6$^\circ$  diameter fully  coded FoV and 
performs  observations  simultaneously with IBIS/ISGRI (fully coded FoV of 9$^\circ$$\times$9$^\circ$) in the energy band 3--35 keV.  Images from JEM--X (3--10 keV and 10--20 keV) were created for all the ScWs during which the source was in the IBIS/ISGRI  FoV from revolution number  906 to 919 (see section 3.1.1), count rates at the position of the source were then extracted from all  individual images to produce the 3--10 keV long-term light  curve on the ScW time scale.

The data reduction was carried out with the release 9.0 of the Offline Scientific Analysis (OSA) software.
Through the paper, the spectral analysis was performed  using \emph{Xspec}  version 12.8.2  and, 
unless stated otherwise,  errors are quoted  at the 90\% confidence level for one single parameter of interest.

\subsection{XMM--Newton} 

For the present study, we analysed a new targeted  {\it XMM-Newton} observation of IGR J18483$-$0311
performed on  2013 April 18. Data reduction was carried out using the latest Science 
Analysis System (SAS v13.5) and following standard procedures. The EPIC-mos (Turner et al. 2001) and EPIC-pn (Struder et al. 2001) 
cameras were operated  in Full Frame mode. The SAS tasks EPCHAIN and EMCHAIN were applied to produce calibrated
event lists for both detectors, respectively. 

The total exposure was  about 58 ks, however after 
subtraction of high flaring background period the net exposure turned out to be 36.2 ks for
the EPIC-pn. 
The source counts were extracted from a circular region of 25" in radius, 
while counts from  the background were extracted from a source-free region in the same CCD.
The redistribution and ancillary matrices were generated using the SAS tasks ARFGEN and 
RMFGEN.  Spectra and light curves were selected from single and double events
only (pattern from 0 to 4) for the EPIC-pn while for both mos cameras patterns from 0 to 
12 were selected. The source net count rates for the three cameras are 0.278$\pm$0.002 (pn), 0.069$\pm$0.001(mos1) and 0.091$\pm$0.001(mos2). The pn has the higher net count rate, hence we decided to consider only data from this 
instrument in the following.  Indeed, the spectroscopy with both the mos cameras does not provide any
improvement with respect to the analysis of the pn spectrum alone. 
Also, the source is too faint for a meaningful spectral analysis with the Reflection Grating 
Spectrometers onboard  {\it XMM-Newton}. The presence of possible pile up was checked using the task EPATPLOT, however no 
significant pile up fraction was found neither in the pn nor in the two mos cameras. Background subtracted light curves were generated 
from event files to which a barycentric correction has been applied using the SAS task BARYCEN. The background count rate was then
subtracted to the light curve with the task EPICLCCORR. In order to guarantee the application of the $\chi$$^2$ statistics, data were grouped to a 
minimum of 25 counts per bin. 

 \begin{figure*}
\begin{center}
\includegraphics[angle=270, width=.50\textwidth]{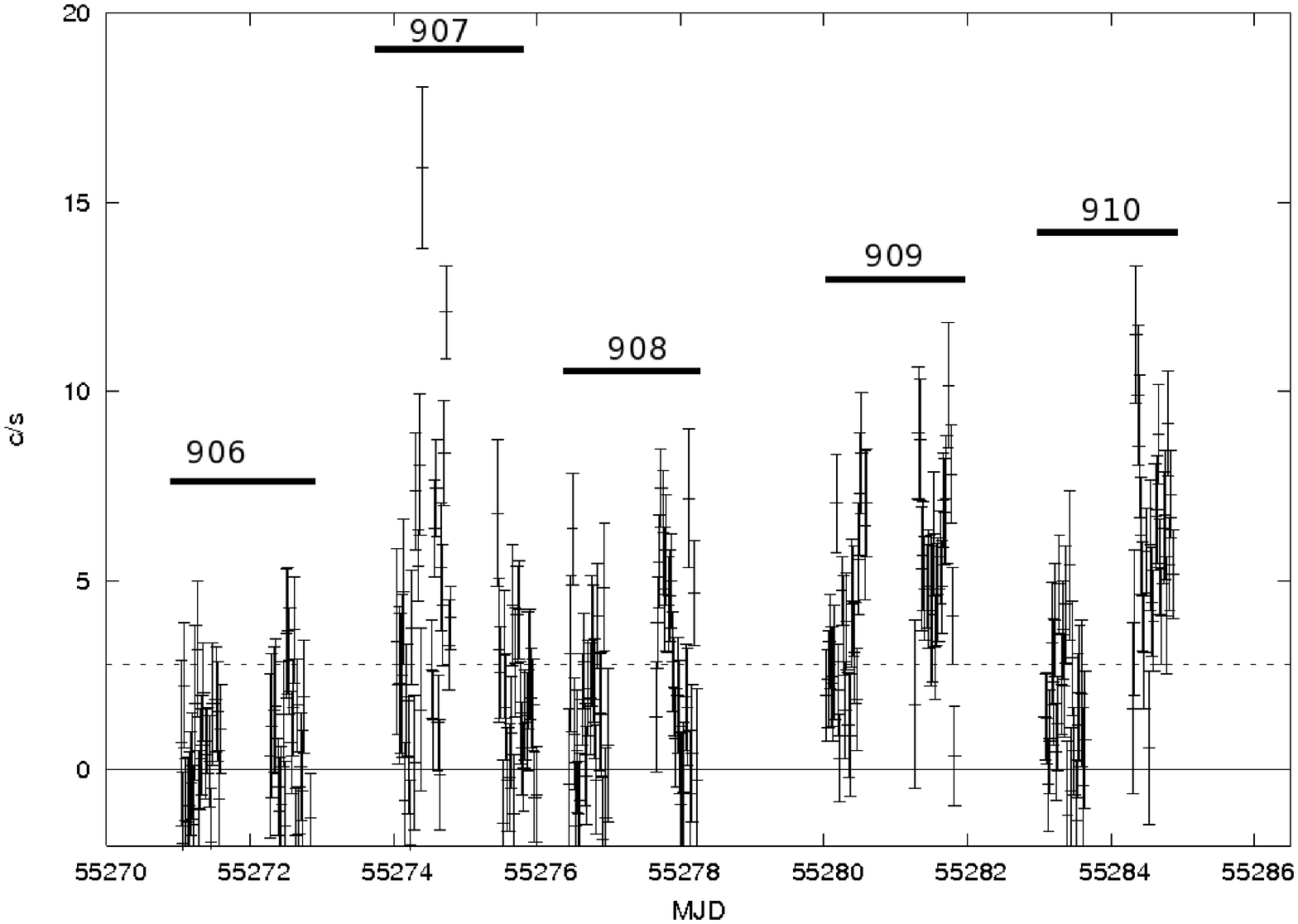}
\includegraphics[angle=270, width=.50\textwidth]{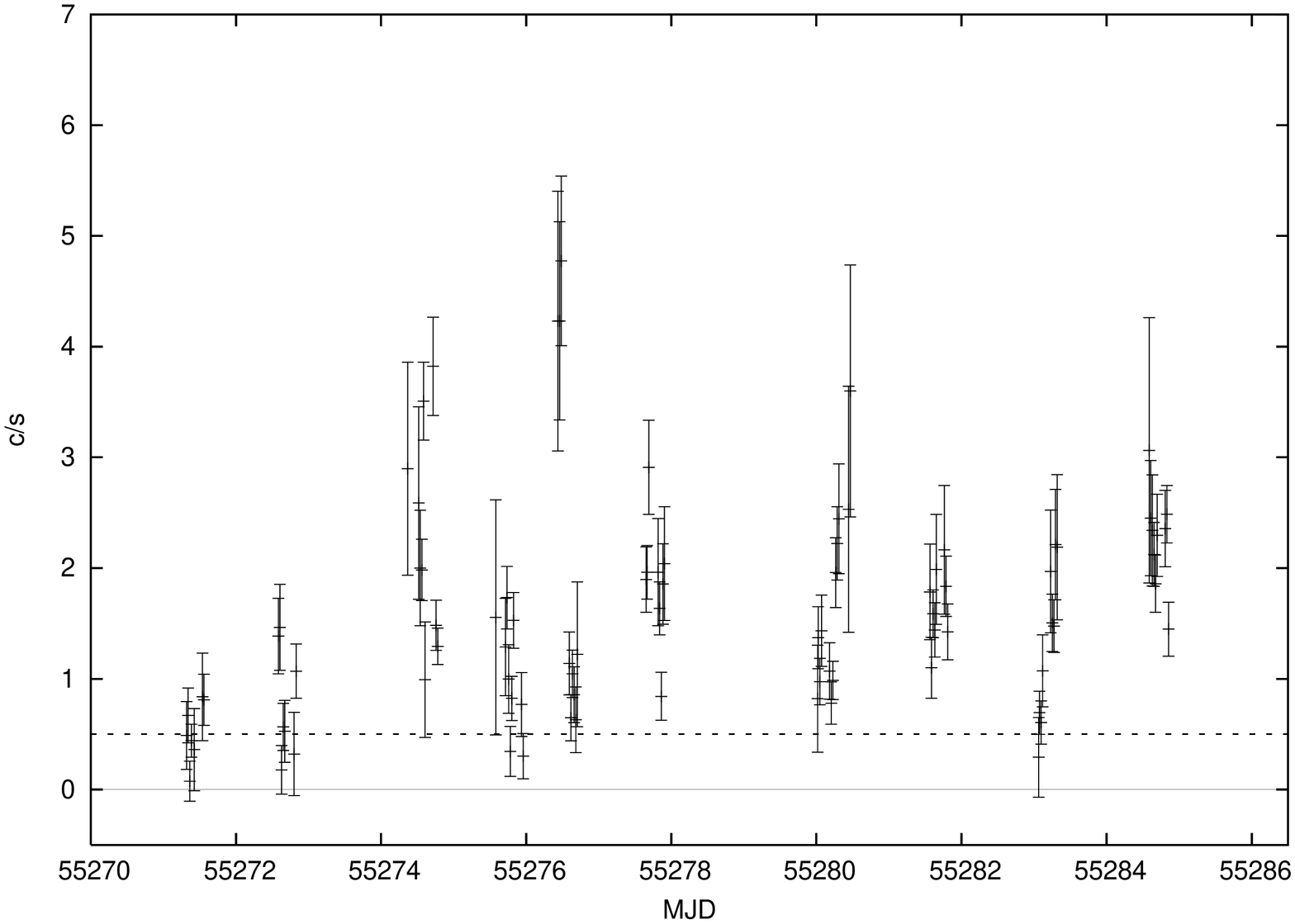}
\caption{IBIS/ISGRI light curve (18--60 keV, top) and INTEGRAL/JEM--X  
light curve (3--10 keV, bottom)  of the  X-ray outburst from IGR~J18483$-$0311. The bin time is 2,000 seconds in both light curves  and 
the dotted black line represents the instrumental 2$\sigma$ upper limit at the ScW level.}
\end{center}
\end{figure*}

\section{INTEGRAL RESULTS}
\subsection{IBIS/ISGRI}
\subsubsection{Unusually long outburst activity}
From an analysis at the ScW level of all the deconvolved IBIS/ISGRI shadowgrams,  we report on a newly discovered and particularly long hard X-ray outburst activity from  the SFXT IGR J18483$-$0311. During the period from  15 March 2010  10:30 UTC to 30 March 2010 02:00 UTC, corresponding to five consecutive INTEGRAL spacecraft revolutions from number 906 to 910,  the source was well covered by IBIS/ISGRI observations being in its FoV  for a significant amount of time.  As we can clearly see in Fig. 1 (top), IGR~J18483$-$0311 was not detected in the 18--60 keV significance image of revolution 906 despite being observed for a total on-source time of 85 ks. On the contrary,  it was significantly detected in all the subsequent four revolutions (from number  907 to 910) with 18--60 keV significance value (exposure) of 15.6$\sigma$ (92 ks), 13$\sigma$ (87 ks),  24.5$\sigma$ (86 ks) and 22.3$\sigma$ (91 ks), respectively. Unfortunately the IBIS/ISGRI temporal coverage of the  source was much  poorer   during all the  subsequent  revolutions  (i.e.  911, 913, 915,  917, 918, 919) being $\sim$ 14 ks exposure in each revolution. No significant detection was obtained in any of these single revolutions, however since such non detections  could   have been largely hampered by the significantly lower exposure times (if compared to those  during previous revolutions)  we summed all the revolutions up   to obtain a  18--60 keV significance image with a relevant total exposure time on-source of $\sim$ 81 ks.  As we can clearly see in Fig. 1 (top), IGR J18483$-$0311 was not significantly detected and so we can safely state that the source was not active anymore from revolution 910 on.   The start time of the outburst activity can be well constrained  at 19 March 2010 UTC 05:13 ($\sim$ 55274.2 MJD), unfortunately its  end time cannot be  constrained  as well because of the much poorer  and  sparser  IBIS/ISGRI temporal coverage of the outburst towards its end of activity. Despite this drawback,  a very conservative end time can be assumed as the finish  of the last revolution (i.e.  910) during which there was highly significant detection (29 March 2010 UTC 21:00).  This implies a conservative lower limit on the outburst duration of the order of $\sim$ 10.6 days.  Table 1 reports the main characteristics of such  X-ray outburst activity as  investigated by INTEGRAL  from revolution 906 to 919.

Fig. 2  (top) shows the IBIS/ISGRI 18--60 keV light curve on ScW timescales (bin time of $\sim$ 2,000 seconds) covering revolutions from 906 to 910,  the dotted black line represents the instrumental 2$\sigma$ upper limit at the ScW level ($\sim $10 mCrab or $1.3\times 10^{-10}$ erg cm$^{-2}$ s$^{-1}$). Clearly, during the entire rev 906 the source was undetected   its flux consistent being with the zero value and well below the instrumental sensitivity of IBIS/ISGRI. Conversely,  in rev 907 the source suddenly  turned on  and flared up reaching a peak flux of 54$\pm$7 mCrab or (7.0$\pm$0.9)$\times$10$^{-10}$ erg cm$^{-2}$ s$^{-1}$ in only $\sim$ 6 hours, the corresponding luminosity is  $\sim$ 7.5$\times$10$^{35}$ erg  s$^{-1}$ (at 3 kpc distance). The dynamic range of the source from the 2$\sigma$ upper limit  in rev 906 (0.9 mCrab)  to the peak-flux is 60. In all the subsequent revolutions from 908 to 910 the source was  active and variable  up to a factor of $\sim$ 5 on ScW timescales.  To investigate possible enhancements of the  variability factor due to more structured and stronger flares  on much shorter timescale,  we produced  the 18--60 keV IBIS/ISGRI light curve with  a finer bin time of 500 seconds,  however no significantly stronger flares where evident in the light curve. 

If we follow Sguera et al. (2007)  and measure the source phase  from the epoch of the brightest outburst observed with INTEGRAL at MJD 53844.1 (phase 0, periastron)  by considering the refined  orbital period value of 18.545$\pm$0.003 (Levine et al. 2011),  then the hard X-ray outburst activity spans orbital phases from $\sim$ 0.91 to $\sim$ 0.45  (i.e.  from close  periastron approach to close apastron approach) and lasted for $\sim$ 60\% of the entire orbital period. 

We extracted the IBIS/ISGRI spectrum of the source from its entire outburst activity (revolution 907 to 910, 38$\sigma$  detection, 355 ks  exposure on-source). The 20--60 keV  spectrum is well fitted by a simple power law model with $\Gamma$=2.8$\pm0.2$ ($\chi^{2}_{\nu}$=1.09, 14 d.o.f.), the average 18--60 keV (60--100 keV) flux  is  equal to  2.1$\times$10$^{-10}$ erg cm$^{-2}$ s$^{-1}$  (3.5$\times$10$^{-11}$ erg cm$^{-2}$ s$^{-1}$) which translates into a X-ray luminosity of 2.3$\times$10$^{35}$ erg  s$^{-1}$   (3.8$\times$10$^{34}$ erg  s$^{-1}$ )   at 3 kpc distance.

For the sake of completeness, we note that Ducci et al. (2013) recently  reported on several outbursts from  IGR~J18483$-$0311 as detected with INTEGRAL in  the energy band  18--50 keV.  In their work,   the authors  mention  the detection of  the source during each single revolution  907, 909 and 910.  However,   such detections were considered  as corresponding to three different and distinct  outburst activities  lasting about  0.5, 1.6 and 0.6 days respectively. 
More importantly,  no  detection of the source in  revolution 908 was reported by  Ducci et al (2013).   We cannot fully establish the origin of the latter difference in results since Ducci et al (2013) did not provide some essential information, specifically the 
significance threshold used for outburst recognition at revolution level.  Without this, or the INTEGRAL significance values (both IBIS/ISGRI and  JEMÐX) pertaining to their detection of the outburst activity from each single revolutions 907 to 910, we cannot explain why they did not detect the signal
in revolution 908. However, we note that revolution 908 is also our lowest flux/significance positive detection in this sequence, so it is possible that it was marginally below their detection threshold while the other revolutions were not.

\begin{figure}
\includegraphics[angle=360, width=.44\textwidth]{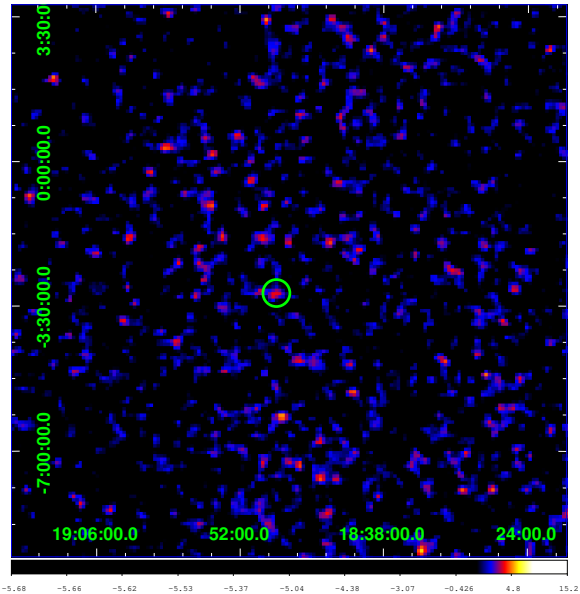}
\caption{IBIS/ISGRI mosaic significance image (100--250 keV). The circle marks the position of  the SFXT IGR J18483$-$0311} 
\end{figure}

\subsubsection{Investigation of the possible hard X-ray tail}

Ducci et al. (2013) recently reported the presence of a possible hard excess at energies above  $\sim$ 80 keV (and possibly extending up to $\sim$ 250 keV) 
in the average 3--250 keV broad band  JEM--X+ISGRI X-ray spectrum of  IGR~J18483$-$0311 in outburst. 
According to the F-test, the putative  hard X-ray tail was  significant at less than 3$\sigma$ level (Ducci et al. 2013). 
It is important to point out that to date no detection of SFXTs at energies above $\sim$ 100 keV has never been reported in the literature, 
both from imaging and spectral analysis.  As a consequence, according to the hint  reported by Ducci et al. (2013), IGR J18483$-$0311 
could possibly be the SFXT with the hardest X-ray spectrum known, with very important implications for the theoretical models 
usually invoked to explain the physical reasons behind the X-ray behavior of SFXTs.  Given the importance of such a point, we performed an 
imaging analysis in the energy band 100--250 keV  to investigate a possible significant detection of the source which could eventually 
confirm or  not the real nature of the putative hard X-ray  tail. Following all the information reported in Table 1 and section 2.1 in Ducci et al. (2013), 
we produced a 100--250 keV mosaic significance image by using all the ScWs during which the source  was significantly 
detected in outburst in the range 18--50 keV (total exposure of $\sim$ 370 ks).  IGR J18483$-$0311 was not significantly detected, with the 
highest significance value (at pixel level) equal to 2.4$\sigma$.  We inferred a 100--250 keV  3$\sigma$ upper limit of  $\sim$ 5 mCrab or 
4.3$\times$10$^{-11}$ erg cm$^{-2}$ s$^{-1}$.  The 100--250 keV mosaic  image  (see Fig. 3)  shows that the sky region is clearly characterized by many  noisy   pixels and/or structures  (due to fluctuations of the background) which have significance values in the  range $\sim$ (2-4)$\sigma$. We conclude that 
the putative hard X-ray excess reported by Ducci et al. (2013) is not real and it is very likely due to noisy  background.

\subsection{JEM--X}

Fig. 1 (bottom)  shows the 3--10 keV JEM--X significance image sequence from revolution 906 to 910.   
Because of  the  much smaller JEM--X   FoV compared to the IBIS/ISGRI one,  the source was inside the 
JEM--X fully coded FoV   during a much smaller number of ScWs and with much lower exposure time. In particular, the source
was never inside the JEM--X  fully coded FoV during all the revolutions after number 910.  
IGR~J18483$-$0311 was barely detected at $\sim$ 2.5$\sigma$ level in the 3--10 keV significance image of revolution 906 (13 ks exposure time), on the contrary   it was significantly detected in all the subsequent four revolutions (from number  907 to 910) with 3--10 keV significance values (exposures) of $\sim$  11$\sigma$ (15 ks), 8$\sigma$ (12 ks),  12$\sigma$ (15 ks), and 11$\sigma$ (12 ks), respectively.  It is worth pointing out  that the source was significantly detected in the higher energy band 10--20 keV as well, with significance values of 8$\sigma$, 5.5$\sigma$, 8.5$\sigma$  and 8$\sigma$, respectively. 

Fig. 2 (bottom)  shows the  3--10 keV JEM--X   light curve on ScW timescales (bin time of $\sim$ 2,000 seconds) covering revolutions from 906 to 910,  the dotted black line represents the instrumental 2$\sigma$ upper limit at the ScW level ($\sim $2 mCrab or $2\times 10^{-11}$ erg cm$^{-2}$ s$^{-1}$). During the entire revolution  906 the source flux value was at the limit of the  instrumental sensitivity of JEM--X. Conversely  in revolution  907 the source suddenly  turned on and  
reached  the highest peak flux of 9$\pm$3  mCrab ((2.8$\pm$0.5)$\times$10$^{-10}$ erg cm$^{-2}$ s$^{-1}$) in revolution 908 (corresponding  to a luminosity of $\sim$ 3$\times$10$^{35}$ erg  s$^{-1}$ at 3 kpc distance). The dynamic range  from the 2$\sigma$ upper limit  in revolution 906  to the peak-flux is about 15  while  when active the source was variable by a factor of $\sim$ 6 on ScW  timescales.  Such JEM--X results clearly confirm that  the SFXT  IGR J18483$-$0311 displayed an unusually long X-ray activity not only above 20 keV but also at lower energies  in the softer X-ray band. 

We extracted the JEM--X spectrum of the source from its entire outburst activity (revolution 907 to 910, 21$\sigma$ detection, 55 ks 
exposure on-source). The 3--20 keV  spectrum is reasonably fitted ($\chi^{2}_{\nu}$=1.8, 29 d.o.f.). 
by an absorbed  power law model with photon index $\Gamma$=1.6$\pm0.2$ and 
intrinsic absorption N$_H$=(6.8$^{+5.1}_{-4.4}$)$\times$10$^{22}$   cm$^{-2}$  in addition to the galactic one along the line 
of sight ($\sim$ 1.6$\times$10$^{22}$   cm$^{-2}$).  The average 3--10 keV (10--20 keV) flux  is  equal to  
1.9$\times$10$^{-10}$ erg cm$^{-2}$ s$^{-1}$  (2.2$\times$10$^{-10}$ erg cm$^{-2}$ s$^{-1}$) which translates into a X-ray luminosity of $\sim$ 
2$\times$10$^{35}$ erg  s$^{-1}$   (2.4$\times$10$^{35}$ erg  s$^{-1}$ )   at 3 kpc distance.


\begin{figure}
\includegraphics[angle=270, width=.45\textwidth]{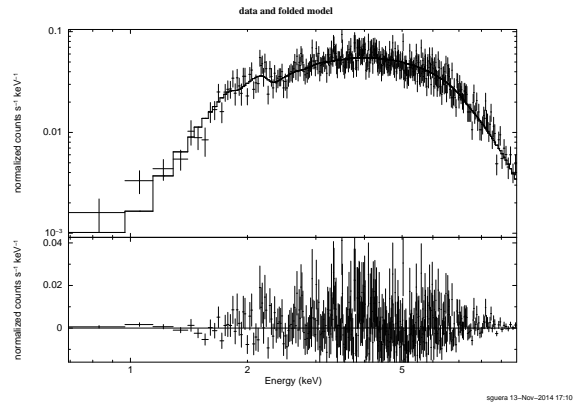}
\caption{EPIC-pn  spectrum of IGR J18483$-$031 best fit with an absorbed black body. } 
\end{figure} 

\begin{figure}
\includegraphics[angle=270, width=.48\textwidth]{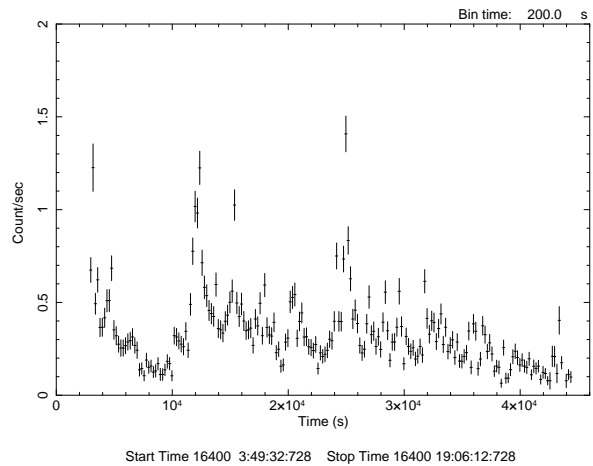}
\caption{EPIC-pn light curve of IGR J18483$-$0311 (0.5--10 keV). } 
\end{figure} 

\begin{figure}
\includegraphics[angle=270, width=.48\textwidth]{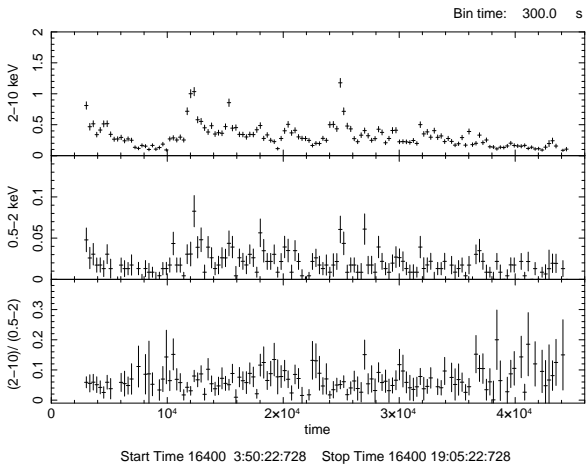}
\caption{EPIC-pn light curves of IGR J18483$-$0311 in the energy bands 2--10 keV (top) and 0.5--2 keV (middle). 
The bottom panel shows the  hardness ratio light curve at the same resolution.} 
\end{figure}

\section{XMM-{\it Newton} RESULTS}

If we follow Sguera et al. (2007)  and measure the source phase  from the epoch of the brightest outburst observed with INTEGRAL at MJD 53844.1 (phase 0, periastron)  by considering the refined  orbital period value of 18.545$\pm$0.003 (Levine et al. 2011),  then the 
{\itshape XMM-Newton} observation took place at orbital phase in the range 0.76--0.80, i.e. during the approach to periastron passage. 

\subsection{Spectral analysis}

The average 0.5--10 keV   EPIC-pn spectrum  was firstly fitted 
with an absorbed power law model ($\chi^{2}_{\nu}$=1.11, 349 d.o.f.)
whose parameter values are intrinsic N$_{H}$=3.7$^{+0.3}_{-0.3}$$\times$10$^{22}$ cm$^{-2}$ (in addition to the galactic one of  1.6
$\times$10$^{22}$ cm$^{-2}$) and $\Gamma$=1.56$\pm$0.08.  The best fit was obtained with an absorbed thermal black body ($\chi^{2}_{\nu}$=1.01, 349 d.o.f., see Fig. 4) which  yielded spectral parameters values of  intrinsic N$_{H}$=0.98$^{+0.16}_{-0.16}$$\times$10$^{22}$ cm$^{-2}$ and 
kT=1.71$^{+0.04}_{-0.04}$ keV.  We note that the  intrinsic N$_{H}$ value is compatible within the errors with the previous  XMM-{\it Newton}  measurement 
(N$_{H}$=1.5$^{+0.6}_{-0.7}$$\times$10$^{22}$ cm$^{-2}$,   Sguera et al. 2010),  on the contrary the black body temperature measurement 
is higher with respect to the previous one (kT=1.35$^{+0.08}_{-0.08}$ keV). This latter difference could be likely explained in terms 
of harder-when-brighter spectral X-ray behavior (as well known in other SFXTs and HMXBs) since  in the present 
XMM-{\it Newton} observation the flux of the source is higher  by a factor of four  with respect to that during the previous observation (Sguera et al. 2010).  The temperature kT resulted  in a radius of the emitting black body region equal to $\sim$0.13 km,  i.e. consistent with a small portion of the neutron star surface such as its polar cap region.   The unabsorbed (intrinsic) 0.5--10 keV flux was 3.8$\times$10$^{-12}$ erg  cm$^{-2}$ s$^{-1}$ (3.2$\times$10$^{-12}$ erg  cm$^{-2}$ s$^{-1}$) which translates into a X-ray luminosity of 4.1$\times$10$^{33}$ erg   s$^{-1}$ (3.4$\times$10$^{33}$ erg   s$^{-1}$)
by assuming a distance of 3 kpc.

\subsection{Timing analysis}

Fig. 5 shows the EPIC-pn background-subtracted light curve in the energy range 0.5--10 keV (bin time 200 s). It is evident that the source displayed variability on 
few minutes timescale by a factor of $\sim$ 15.  Flickering activity occurred  sporadically,   following time intervals of lower X-ray emission.  In Fig. 6  we report the  light curves  in the softer (0.5--2 keV) and harder (2--10 keV) X-ray bands, in particular  the lower panel 
shows  the hardness ratio which  clearly indicate that there  is no appreciable and significant spectral variation. 

We searched for the $\sim$ 21 seconds  spin period of the neutron star hosted in IGRJ18483$-$ 0311 (Sguera et al. 2007, Giunta et al. 2009).
We applied the Solar System barycenter correction to the photon arrival times with the SAS task barycen.  Unfortunately, due to the particularly  weak signal  and  low source flux,  the production of finely binned  light  curves (e.g.  1 s,  4 s) resulted in  flux values consistent with zero in a lot of the data points in the light curve, preventing a proper  and thorough testing of the $\sim$ 21 s periodicity.  Consecutively, barycentred light curves  were extracted with  larger time resolution of  8 s and 10 s in several energy bands (e.g. 0.5--10 keV, 2--10 keV, 0.5--5 keV, 5--10 keV)  and tested for periodicities  via the Lomb--Scargle method by means of the fast implementation of Press \& Rybicki (1989) and Scargle (1982).  Periodicities were searched  in the frequency range from  0.000047 Hz  (after which the sensitivity is reduced due to the finite length of the light curves) to 0.06 Hz or 0.05 Hz (corresponding to the Nyquist frequency of the data set),  however the intrinsic uncertainties of the light curve data points resulted in a noise-dominated periodogram without  
any significant and unambiguous evidence for coherent modulation. 
Since the Lomb-Scargle method is generally preferred for data set with gaps and unequal sampling (which is not the case of the present 
XMM--{\it Newton} observation),  after barycentric correction of the photon arrival times in the original event lists  we performed a Fast Fourier Transform analysis.  A  power spectrum was produced covering the frequency interval as above, however no significant evidence for a peak  was found as well. The fractional amplitude to which we are  sensitive can be calculated according to Luna \& Sokoloski (2007) and it   was found that we
are sensitive to oscillations with fractional amplitudes of  $\sim$ 8\%  (0.5--10 keV).  
As a subsequent step, we performed an epoch folding analysis. Periodicities were searched for in a small frequency window centred 
at the putative period of 21 seconds. While a large value of $\chi^{2}$ would represent a robust indication of a periodic modulation, again no peak with a significantly high value of $\chi^{2}$  was found. We note that the strongest peak  ($\chi^{2}$=36.2, 7 d.o.f.) in the  $\chi^{2}$ vs P$_{trial}$  plot
corresponded to a best period of $\sim$ 21.03  
s,  but was statistically significant  at only  $\sim$ 3 sigma level (corrected for the number of  trial periods, 150).

 \section{DISCUSSION AND SUMMARY}

We  presented  a  broad band X-ray study (0.5--250 keV) of the  SFXT IGR J18483$-$0311 
using archival INTEGRAL data and a new targeted {\itshape XMM-Newton} observation.
 
Our INTEGRAL investigation discovered for the first time a particularly  long  X-ray activity (3--60 keV)  which continuously lasted  for
at least $\sim$ 11 days, i.e. a very significant fraction ($\sim$ 60\%) of the entire orbital period.  
This characteristic is  at odds with the much shorter durations marking  outbursts from classical SFXTs especially above 20 keV
both in term of absolute time (typically from few hours to very few days )  and fraction  of the orbital period (typically $\le$ 20\%).    
It is worth pointing out that previous works (Sguera et al. 2007, Ducci et al. 2013) already reported on some X-ray outbursts from  the SFXT IGR~J18483$-$0311 lasting a few days.  As useful example, Fig. 7 shows the IBIS/ISGRI light curves of the two previously discovered and longest outbursts having  duration in the  range  2--3 days (see details in Sguera et al. 2007).  Remarkably,  the  newly discovered X-ray activity  reported in this work lasted significantly  longer (at least $\sim$ 11 days) and  this  adds a further extreme characteristic to the already extreme   SFXT IGR J18483$-$0311.  
For the sake of completeness,  we note that the source  IGR J11215$-$5952  is the only other  SFXT known to be also characterized  by  X-ray outburst activity with a similarly long duration of  $\sim$ 6--8 days (Romano et al. 2009). However,  as recently highlighted by Lorenzo et al. (2014),  
IGR J11215$-$5952  is a peculiar SFXT with very  eccentric and unusually long orbital period of $\sim$ 165 days,  such  characteristics (as well as its 
strictly recurrent X-ray behavior and its position in the Corbet diagram) are akin to those of transient Be HMXBs. Contrarily to other classical SFXTs, during its unusually long X-ray activity IGR~J18483$-$0311 was not characterized by fast/strong X-ray flares but    it rather displayed  an enhanced X-ray flux  moderately  variable by  a factor of $\sim$ 5 both in the hard and soft X-ray bands.  At the peak the 18--60 keV  (3--10 keV) luminosity was  equal to 7.5$\times$10$^{35}$ erg  s$^{-1}$ (3$\times$10$^{35}$ erg  s$^{-1}$)  i.e. the source was under-luminous by a factor of a few relative  to its  typical outbursts L$_{x}$  of the order of $\sim$ 10$^{36}$ erg  s$^{-1}$ (Sguera et al. 2007).  The particularly long X-ray activity detected by INTEGRAL  lasted for a significant fraction ($\sim$ 60\%) of the entire $\sim$ 18 days orbital period ,  spanning from  orbital phase  $\sim$  0.91 (very close  to periastron  approach at 0 phase) to $\sim$ 0.45 (very close to apastron approach at 0.5 phase). We note that   previous  INTEGRAL and  {\itshape XMM-Newton}  investigations around apastron (Sguera et al. 2007, Sguera et al. 2010) showed that the source was predominantly undetected at hard X-rays (L$_x$$<$10$^{34}$  erg  s$^{-1}$) and  very weak at soft X-rays (L$_{x}$ $\sim$ 1.3$\times$10$^{33}$ erg  s$^{-1}$). On the contrary the present study reports for the first time relevant  X-ray activity during apastron,  with X-ray  luminosity values more  than two orders of magnitude greater with respect to previous measurements both above and below 10 keV. From IBIS/ISGRI and  JEM--X analysis, the  X-ray spectral shape was identical to that of  many other  X-ray outbursts already reported in the literature  (Ducci et al. 2013, Sguera et al. 2007). 

\begin{figure}
\includegraphics[width=.43\textwidth]{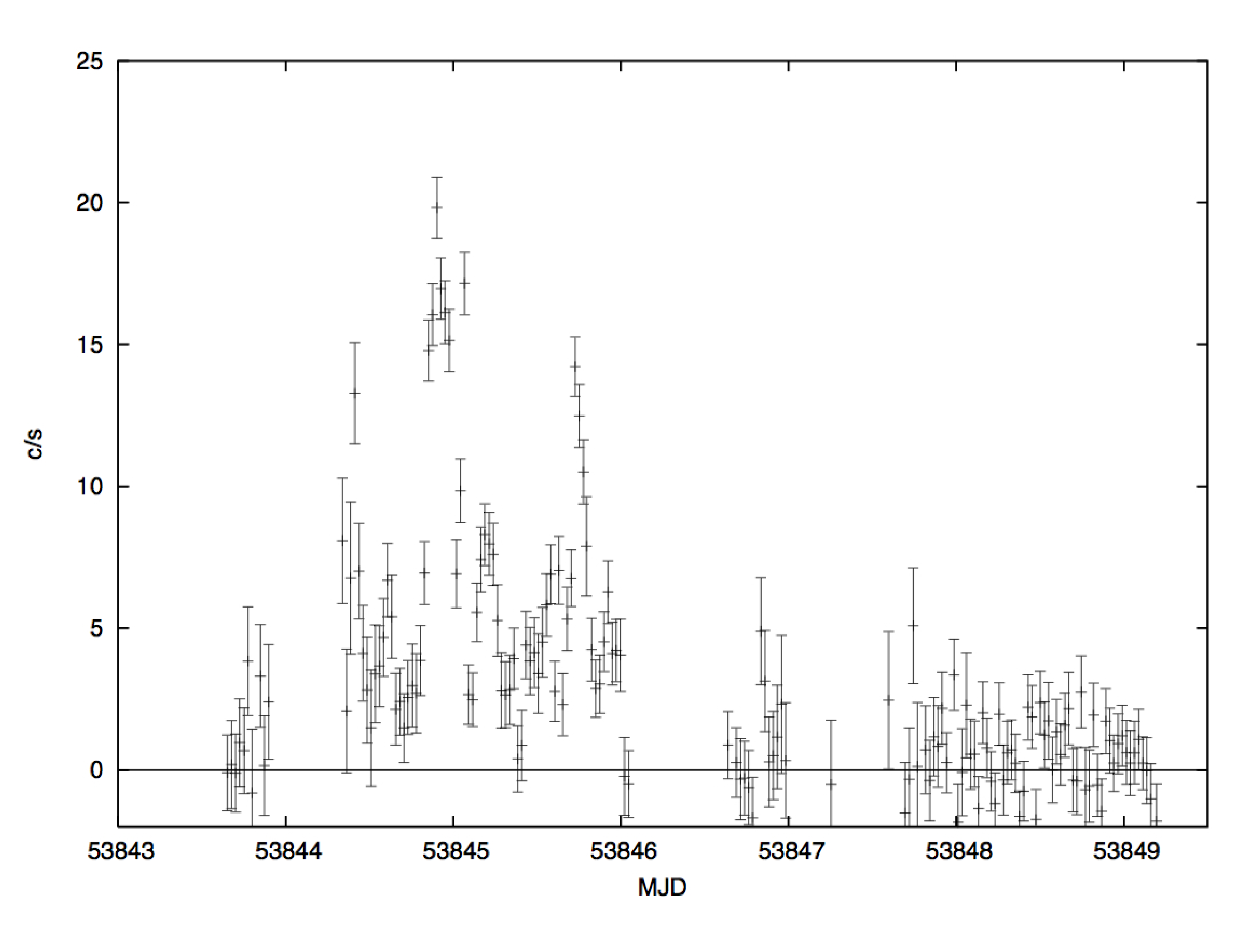}
\includegraphics[width=.43\textwidth]{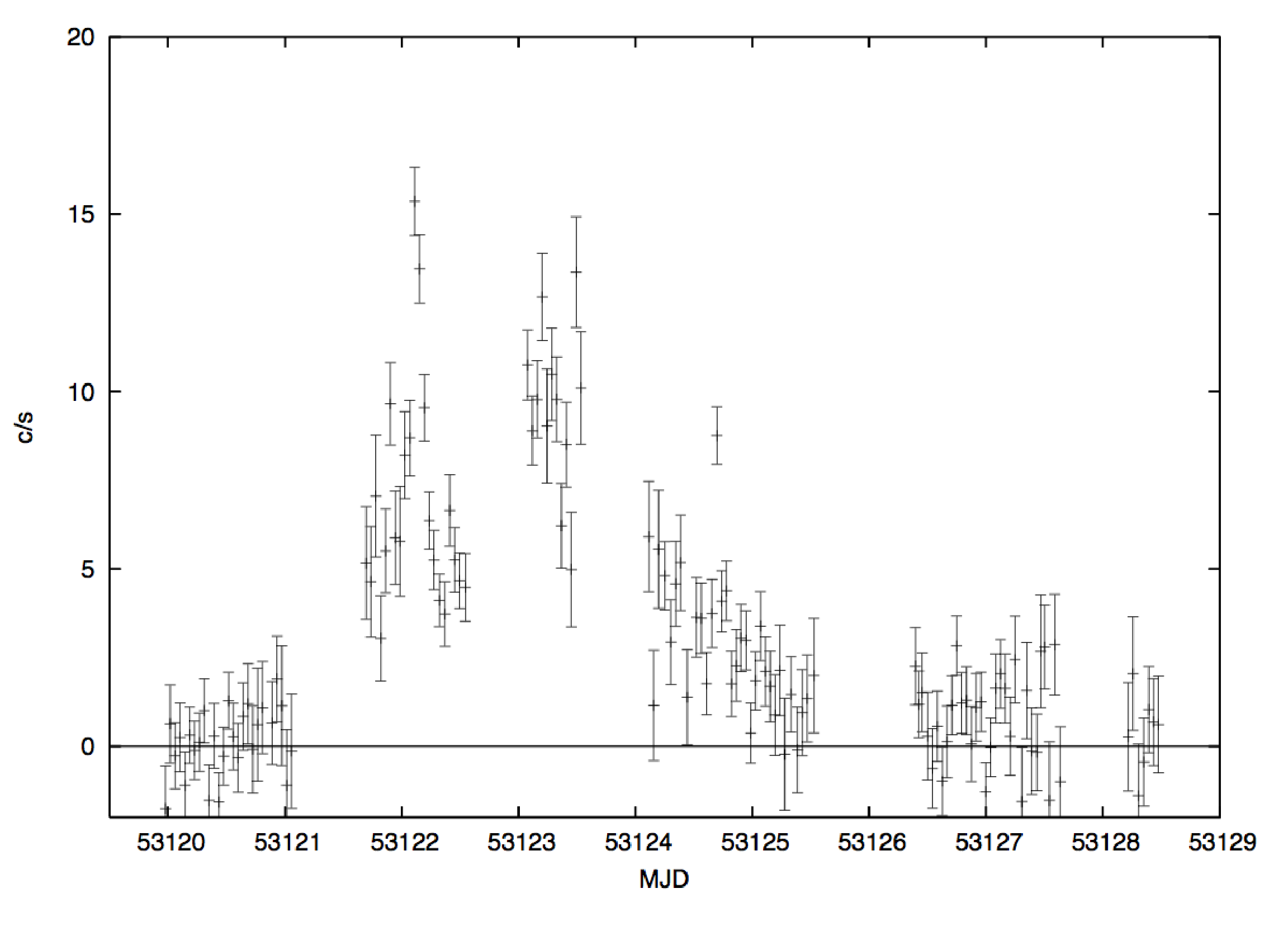}
\caption{IBIS/ISGRI light curves (18--60 keV) on ScW timescales (2,000 seconds) of the two  longest previously reported X-ray 
outbursts from IGR~J18483$-$0311.} 
\end{figure}

From a  new {\it XMM-Newton} targeted observation,  we reported  the spectral and temporal properties of the source. 
Although the  observation  took place  at periastron approach, IGR~J18483$-$0311 was detected at low X-ray luminosity ($\sim$  4$\times$10$^{33}$ erg   s$^{-1}$, 0.5--10 keV) showing variability by a factor of $\sim$ 15 on few minutes timescale. This kind of  behavior at low L$_x$ has already been observed in other SFXTs  (e.g. Bozzo et al. 2010, Sidoli et al. 2010),  it is well known that SFXTs  spend the great majority of the time in  this so-called intermediate X-ray state 
(Sidoli et al. 2008) with typical hard X-ray spectra ($\Gamma$$\sim$ 1--2) as result of  accretion of  material  at a much lower rate   than that during the bright X-ray outbursts .  The hard X-ray spectrum and the low X-ray flux  measured by   {\it XMM-Newton}  are compatible   with this intermediate intensity state scenario.  
The EPIC-pn spectrum is best fitted by an absorbed blackbody,  in particular  we note that, contrarily to Sguera et al. (2010),  no emission line feature at $\sim$ 3.3 keV  was evident  in the spectrum of the present new {\it XMM-Newton} observation. Such a line  was previously interpreted by Sguera et al. (2010) in term of cyclotron emission feature following the theoretical  framework of  Nelson et al. (1993, 1995), and implies  a magnetic field value of about  4$\times$10$^{11}$ G with very important implications on theoretical models. In the light of the newly reported  results,  the real nature of the putative cyclotron line is  called into question  and  it  should be treated with caution. However we note that according to the theoretical framework  of Nelson et al. (1993,1995) this kind of cyclotron line,  if present,  becomes detectable only during the  lowest  X-ray luminosity state of transient HMXBs when their X-ray continuum is the weakest possible.  Unfortunately  during the present  new {\it XMM-Newton} observation (around periastron passage)  the X-ray luminosity  of the source   was  higher  by a factor of four  with respect to that during which the putative line was previously  detected   (around apastron passage) at the lowest ever measured X-ray luminosity (Sguera et al. 2010) . It is likely that  the line (especially if weak)  was  not detectable in the present  new {\it XMM-Newton} observation because it was 
drowned out  by the  brighter continuum. Only detailed studies  performed at  the similarly lowest X-ray luminosity states  (achievable only during the apastron passage) can draw  a definitive judgement on the nature  of the putative line. The  EPIC-pn light curves were  searched   for  the $\sim$ 21 s  spin period of the neutron star hosted in IGR~J18483$-$0311. No significant periodicity  was found by performing a Lomb-Scargle and fast Fourier  transform analysis and we  inferred a fractional amplitude  to which the {\it XMM-Newton} observation is 
sensitive equal to $\sim$ 8\%.  By performing an epoch folding analysis,  the strongest peak  in the $\chi^{2}$ vs P$_{trial}$ plot  corresponded to a best period of $\sim$ 21.03 seconds, however it  was statistically significant  at only  $\sim$ 3 sigma level. 
We caution that the poor statistical quality of the data did not permit us to obtain meaningful high time resolution light curves  (e.g. 1 s or 4 s bin time)  
which are required   for  a detailed timing analysis.  In addition, we note that  the  measured pulsed fraction  reported  by Giunta et al.  (2009)  in a similar energy band  and  at  similarly  low flux level during a different {\it XMM-Newton} observation is particularly low, i.e. in  the range 12\%-18\%.  Our  inferred {\it XMM-Newton} fractional amplitude ($\sim$ 8\%) is notably close to such range of values, it cannot  be excluded  that  even a very small variation and decrement  of the pulsed fraction (which is likely to happen) could have been sufficient  to hamper its significant detection.  
In this respect, it is worth pointing out that when  the $\sim$ 21 s coherent pulsations were discovered in the  JEM--X data by using the Lomb-Scargle analysis (Sguera et al. 2007), the  source was in a very bright outburst activity with a 3--10 keV   L$_x$ of $\sim$ 2$\times$10$^{37}$ erg   s$^{-1}$, i.e. almost four orders of magnitude brighter than the present  {\it XMM--Newton} observation. Recently Ducci et al. (2013) called into question such $\sim$ 21 s pulsation by reanalyzing the JEM--X outburst data,  however here we note that, among the others,  the authors did not completely reproduce the same data analysis as in Sguera et al. (2007),  e.g.  the  temporal range of their analyzed outburst (MJD 53844.6--53846.0)  is  shorter (by $\sim$ 0.5 days)  with respect to that  analyzed by Sguera et al. (2007). It is  possible that in their  timing analysis Ducci et al. (2013) missed a significantly bright portion of  the outburst activity as can be clearly seen from the corresponding light curve reported in Sguera et al. (2007). 

Previous works (Rahoui \& Chaty 2008, Romano et al. 2010) explained the observed X-ray outbursts   from the SFXT IGR~J18483$-$0311 by  
using the spherically symmetric clumpy wind scenario. Within this theoretical framework, it is generally assumed that  the source goes into X-ray outburst activity  preferentially   during the  periastron passage when orbiting  inside regions of the supergiant wind characterized by  a very high density of clumps much more massive than the background wind. In particular,  Rahoui \& Chaty  (2008) found that a very eccentric orbit ({\it e}=0.4--0.7)  is strongly needed to explain the  few days duration of some  observed X-ray outbursts (as those reported in Sguera et al. 2007).  In addition, assuming a large  eccentricity of {\it e}=0.4 Romano et al. (2009) explained the observed X-ray outbursts  and their pronounced short time scale variability  in terms of accretion of single massive clumps composing the donor wind and having masses in the range 10$^{18-20}$ g.   We point out that the following   new findings reported in the present study are hardly  explainable  by  such spherically symmetric  clumpy wind scenario: i)  both the significantly long X-ray outburst duration (at least $\sim$ 11 days) and the small variability factor ($\sim$ 5)  observed by INTEGRAL  cannot be naturally explained  by the accretion of single massive clumps  embedded in a much less dense background wind. An unusually  large number of   clumps,  incessantly accreted, would be required  to continuously  sustain the  prolongated accretion of material  for at least $\sim$ 11 days uninterruptedly ii)  part of the X-ray outburst activity was  detected by INTEGRAL  during the  apastron approach, i.e. right  when  the neutron star  is supposed to be incapable
of accreting material because well outside the region of high density clumps which is encountered  only around periastron passage. The measured X-ray outburst luminosity, in the range (3--7)$\times$10$^{35}$ erg   s$^{-1}$,  is too high to be explained in terms of  accretion from the much less dense background wind iii)  the very low X-ray luminosity of $\sim$  4$\times$10$^{33}$ erg   s$^{-1}$ was measured  by {\it XMM-Newton}  during the approach of the periastron passage  when it is normally expected that the source goes into X-ray  outburst.  This missing  activity is usually explained  in term of extremely varying stellar wind environment (e.g. density and number of clumps) as probed by the neutron star on short orbital period time scale ($\sim$ 18.5 days in our specific case),  which is  somehow  very difficult to  accommodate and explain. On the contrary,  the newly reported  results  have less shortcomings if explained  within the quasi-spherical settling accretion model. In fact,  i) the low X-ray luminosity  measured by   {\it XMM--Newton}  is well below the critical value of $\sim$ 3$\times$10$^{35}$ erg   s$^{-1}$,  this means that the source was inhabiting the  radiatively cooled regime which allows only very low accretion rate onto the neutron star and so low L$_x$ values, independently from the orbital phases of the  system ii) the  outburst X-ray luminosity measured by INTEGRAL is  above the  critical value  $\sim$ 3$\times$10$^{35}$ erg   s$^{-1}$,  and  so the source  was inhabiting   the Compton cooling dominated  regime which allows  higher  accretion rate onto the neutron star leading to the production of  moderately bright flares with L$_{x}$ $\leq$10$^{36}$ erg   s$^{-1}$ as predicted by Shakura et al. (2014).  The  transition between the two regimes may be due to a switch in the X-ray beam pattern in response to a change in the optical depth in the accretion column with varying luminosity (Shakura et al. 2012, 2013).  The triggering for the switch  may be simply due to even very small and modest changes of the local wind velocity and density values,  a condition  very likely  satisfied in a  moderately clumpy wind scenario which realistically invoke  smaller and lighter clumps, not necessarily very massive and large clumps as in the classical spherically symmetric clumpy wind scenario. Notably, the conditions for the transition are independent from the orbital phases of the system and so are not necessarily linked to periastron or apastron passages.

 We note that the peculiar place occupied by the source in the Corbet diagram, lying close to the typical location of the Be HMXBs, lead Liu et al. (2011) to suggest that  IGR~J18483$-$0311  could be the descendant of a Be HMXB system. Liu et al. (2011) discuss this system as a source that is presently a wind-fed system, but it is tempting to suggest, given the longer outburst we discovered, that a variability in the donor wind conditions could have triggered the temporary formation of an accretion disc near the periastron passage, sustaining an unusually longer outburst due to the longer viscous disc timescale. Variable wind conditions at the neutron star orbit could have been due to the crossing of a higher density large-scale wind variable structure similar to the spiral structures/shells discussed by  Lobel et al. (2008). The different wind properties could have played a role in triggering an enhanced and longer accretion phase also within the quasi-spherical wind accretion scenario.  We note that also another SGXB accreting pulsar with a peculiar location in the Corbet diagram, OAO1657$-$415, has been suggested to sometime undergo disc accretion phases, given its long term secular spin-up of its pulsar (Mason et al. 2009). Interestingly enough, also OAO1657-415 has been suggested to be a SGXB evolved from a Be HMXB (Liu et al. 2010). 
 Unfortunately, we cannot say anything conclusive about this scenario, given the lack of a long-term history of the IGRJ18483$-$0311 spin period 
 derivative. 
 
The newly discovered  and unusually long X-ray activity observed from the SFXT IGR~J18483$-$0311
represents a departure from the nominal behavior of classical SFXTs with their much shorter outbursts and it adds a further extreme 
characteristic to the already extreme SFXT  IGR J18483$-$0311. Further INTEGRAL studies on the classical SFXTs searching for similarly unusually long X-ray outbursts are needed to understand if the reported peculiar characteristic observed from  IGRJ18483$-$0311 is exceptional or not among the class of SFXTs.

\section*{Acknowledgments} 
italian authors acknowledge the ASI financial support via grant ASI/INAF n. 2013-025.R.O.

\end{document}